\documentclass[journal,10pt]{IEEEtran}
\usepackage{color}
\usepackage{booktabs}
\usepackage{threeparttable}
\definecolor{gray}{rgb}{0.6,0.6,0.6}

\usepackage{cite}

\ifCLASSINFOpdf
\usepackage[pdftex]{graphicx}
\graphicspath{./Figure/}
 \DeclareGraphicsExtensions{.pdf,.jpeg,.png,.eps}
\else
\fi
\usepackage{amsmath}
\usepackage{amssymb}

\usepackage[caption=false,font=footnotesize]{subfig}
\begin{document}
	%
	\title{Topology Virtualization and Dynamics Shielding Method for LEO Satellite Networks}
	%
	%
	%
	
	
	\author{Quan Chen,
		Jianming Guo,
		Lei Yang,
		Xianfeng Liu
		and~Xiaoqian Chen
		\thanks{This work was supported by National Key Research and Development Program of China, under Grant 2016YFB0502402.}
		
		\thanks{Q. Chen, J. Guo, L. Yang and X. Liu are with the College of Aerospace Science and Engineering, National University of Defense Technology, Changsha, 410072, China (e-mail: chenquan11@foxmail.com; gjm08110@hotmail.com; craftyang@163.com; liuxianfeng\_edu@163.com).}
		\thanks{X. Chen is with the National Innovation Institute of Defense Technology, Chinese Academy of Military Science, Beijing, 100071, China (e-mail: chenxiaoqian@nudt.edu.cn).}
	}
	
	%
	%

	\markboth{IEEE COMMUNICATIONS LETTERS,~Vol.~0, No.~0, 2019}%
	{Shell \MakeLowercase{\textit{et al.}}: Bare Demo of IEEEtran.cls for IEEE Journals}
	%



	\maketitle
	
	\begin{abstract}
		Virtual Node (VN) method is widely adopted to handle satellite network topological dynamics. However, conventional VN method is insufficient when earth rotation and inter-plane phase difference are considered. An improved VN method based on Celestial Sphere Division is proposed to overcome the defects of the conventional method. An optimized inter-satellite link connecting mode is derived to achieve maximal available links. The optimal VN division solution and addressing scheme are designed to generate a nearly static virtual network and solve the asynchronous switches caused by inter-plane phase difference. Comparison results demonstrate the advantages of proposed method.
	\end{abstract}
	

	%
	\IEEEpeerreviewmaketitle

	\section{Introduction}
	
	\IEEEPARstart{t}{he} high-speed movement of low-Earth orbit (LEO) satellites causes drastic changes in satellite network topology. The topological dynamics brings severe difficulties for the network management and routing protocol design, especially in recent popular mega-constellation networks \cite{YangXin}. For widely adopted polar orbit constellation networks, inter-satellite link (ISL) may experience frequent on-off switches \cite{Cola, LuYong2013}. The seam and earth rotation also aggravate topological dynamics.
	
	Many topological dynamics shielding methods have been proposed and can be classified into two types: Virtual Topology (VT) \cite{JiaMin2017, YangXin,Ruiz} and Virtual Node (VN) methods. VN method divides the earth surface into several cells called VN and maps satellites to VNs. In the generated virtual network, topological dynamics is handled. VN method is developed and widely applied to satellite network routing algorithms \cite{Ekici2001, LiuZhiguo2019, ChenQuan2019}.
	
	However, the topological dynamics has not been completely shielded by conventional VN method. Firstly it neglects the earth rotation which makes VN updates within orbit plane cannot support long-term coverage of fixed ground cell. The seam blocking effect\cite{Ruiz} and inter-plane phase difference may also introduce connectivity changes to the virtual network\cite{LuYong2013}. Moreover, the application of conventional VN method requires antenna supporting Earth-fixed mode \cite{Korcak2009}, adding  complexity for the satellite platform and causing severe call dropping problems when switches occur.

	To overcome the defects of conventional VN method, we develop an improved VN method based on celestial sphere division. Since the orbit planes are relative stationary to the celestial sphere, this VN division is independent of earth rotation and the seam location is fixed, while Earth-fixed mode is not required. When inter-plane phase difference is considered, we derive the optimized inter-satellite link (ISL) connecting mode and modify the VN method. Finally, the performances of the two VN methods are compared. The results show the improvement of the proposed method in topological dynamics shielding and ISL availability.
	
	

	\section{Network Topology model}
	\subsection{Satellite Constellation Architecture}
	
	This paper focuses on polar orbit (also named Walker-star) constellation which is widely adopted by existing and newly proposed systems. $ N_S $ satellites are evenly distributed in $ n_1 $ planes with $ n_2 $ satellites in each plane. All the LEO satellites have the same altitude and near 90 deg inclination angle. The constellation is $\pi$-type and the angle between two adjacent planes is $\Delta \Omega =\pi /{{n}_{1}}$. The angle between two adjacent satellites within plane is $\omega_f =2\pi /{{n}_{2}}$.  
	Generally a satellite is equipped with four ISLs, two intra-plane ISLs (named V-ISLs) connecting satellites in the same plane and two inter-plane ISLs (named H-ISLs) connecting satellites in adjacent planes. Some disconnections in specific regions may occur, 
	{which will be discussed in the following section.}
	
	\begin{figure}[t!]
		\centering
		\includegraphics[width=0.8\linewidth]{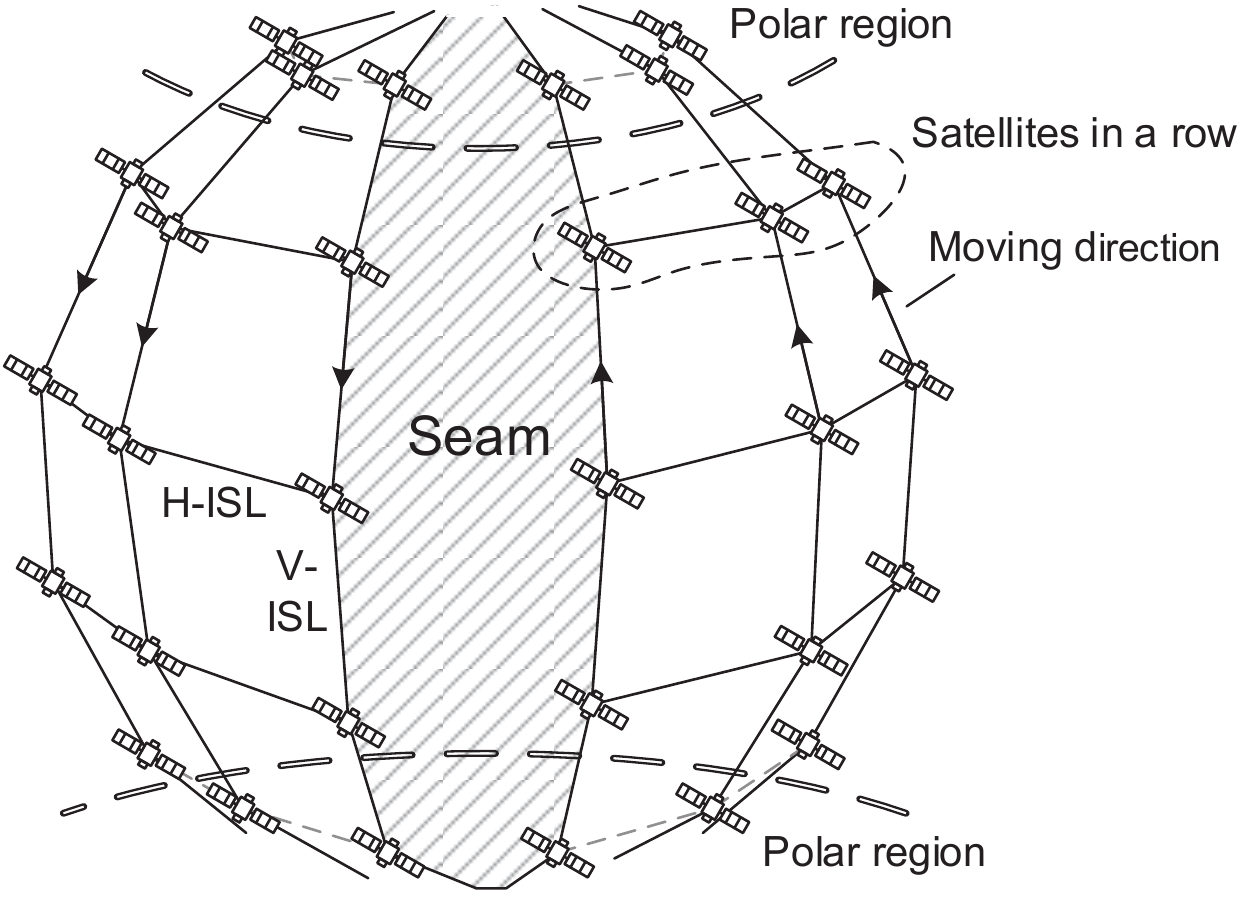}
		\caption{Satellite network topology and ISLs.}
		\label{fig:constTopo}
	\end{figure}

	\subsection{Network Topological Dynamic Characterizations}
	Due to the high-speed relative motion, although satellites are regularly distributed in the constellation, the network topology (see Fig. \ref{fig:constTopo}) is constantly changing. From the view of abstract network topology, the physical distance variation of links can be neglected, while the topological dynamics is mainly reflected on the ISL on-off switches. 
	\subsubsection{On-off switch when flying over polar regions}
	Orbit planes intersect at polar regions. The direction of H-ISLs changes violently when satellite approaches the pole, causing the increase of antenna tracking cost and link instability. 
	Thus, H-ISLs are temporarily shut off when satellites fly into polar regions and re-established when satellites fly out.
	
	\subsubsection{No inter-plane ISL across the seam}
	Satellites in the first and $n_1$-th plane move in opposite directions so that H-ISLs between these two planes can hardly be maintained. Here H-ISL is not established, which generates two "seams" separated by 180 deg.  Satellites beside the seam have to make a detour to access satellites on the other side. 
	
	\section{Conventional Virtual Node Method}
	Satellite network dynamics can be shielded by topology virtualization. In conventional VN method, by dividing the earth surface into several grids, each grid is regarded as a VN and assigned with a unique virtual address. Satellite antenna requires Earth-fixed mode supporting so that it can keep pointing to the fixed grid during the satellite movement. When satellite flies away, the next satellite in the same plane takes over the VN and inherits the network states of the predecessor\cite{Ekici2001}. Each satellite covering the grid is mapped to the VN address. By dynamically mapping satellites to corresponding VNs, the dynamic satellite topology is transformed to a static virtual topology that facilitates the routing issues. 
	{The switching interval can be calculated by 
		\begin{equation}
		T_{S-GRD}=T/n_2\,,
		\end{equation}
		where $T$ is the satellite orbital period\cite{LuYong2013}.}
	
	In this paper, the above-mentioned VN method is named Geographic Region Division based VN method (GRD-VN). GRD-VN can effectively solve the ground user mobility management and polar region switches. However, the generated virtual network is not completely static in practice. 
	
\begin{figure}[t!]
		\centering
		\includegraphics[width=\linewidth]{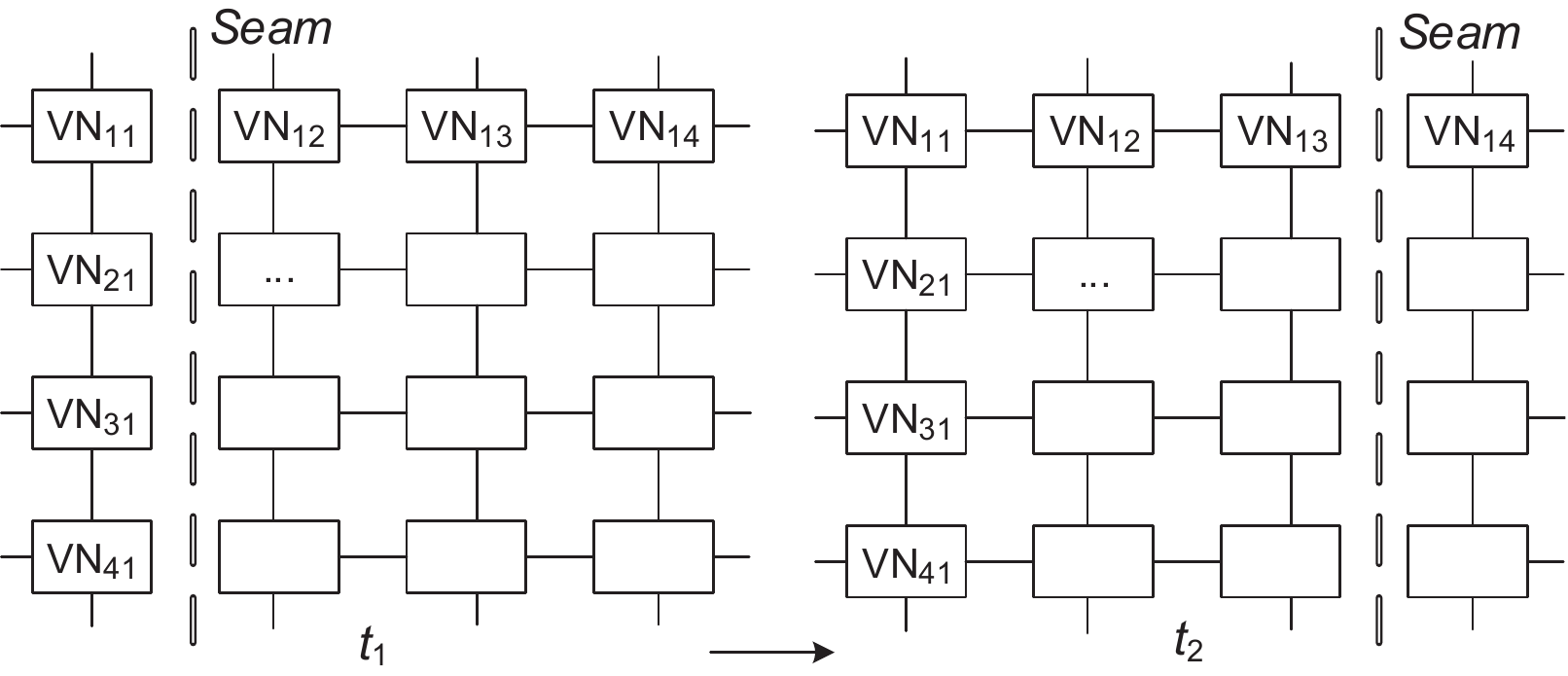}
		\caption{Seam movement and blocking effect in GRD-VN virtual network.}
		\label{fig:GRD_VN_Seam}
	\end{figure}

	The key problem is that it neglects the earth rotation. When initial VN moves far from the corresponding orbit plane, satellites are unable to support VN coverage \cite{LuYong2013}. Although the earth rotation effect can be partly avoided by assigning VN to satellites in  adjacent orbit\cite{Korcak2009}, the topology dynamics caused by the seam becomes a new challenge. If a VN is served by satellites of different orbits, the seam location in the virtual network cannot be determined but moves with the earth rotation (see Fig. \ref{fig:GRD_VN_Seam}). Since the H-ISLs across the seam are not established, links between the virtual nodes are not permanent.
	
	Furthermore, conventional VN method does not consider the switching synchronization issues caused by satellite phase difference. Satellites in different planes connected by H-ISLs are defined as satellites in a row. When satellite phase difference exists, "asynchronous switches" in a row may happen because some satellites fly into polar regions first while others follow behind. Then VN addresses in this row are disrupted and it is difficult to determine VN address from the geographic location. 

	\section{Proposed CSD-VN method}
	To overcome the deficiency of GRD-VN, we propose an improved version -- Celestial Sphere Division based Virtual Node (CSD-VN) method. Celestial sphere is an abstract sphere used in astronomy to demonstrate location of planets or satellites. By dividing the celestial sphere into grids, each celestial grid is regarded as a Virtual Node with a virtual address and is served by the satellite in the celestial grid. Satellites are mapped to celestial grids and share the corresponding virtual address.
	
	The VN virtual address is a two-element array denoted by $(v,h)$ indicating
	the $v$-th satellite in the $h$-th orbit plane. When the satellite moves, the $h$-index keeps unchanged while $v$-index is updated with the satellite latitude. Thus, the VN is bound to specific celestial region and in accordance with satellite distribution. By appropriate regional division, the VN in the polar region and beside the seam can be pre-determined and kept constant. Consequently, in the space segment the virtual network becomes static during the system operation.
	
	\subsection{Without inter-plane phase difference}
	According to the definition, satellites in a row have the same $v$-index. If the phase difference of satellites in adjacent orbits $ \Delta f $ is zero, then satellites in a row have the same latitude, and the switches at the polar region border are synchronous.
	
	{The VN is a one-to-one mapping from the virtual network to the satellite set $S$.}  $h$-index is numbered from west to east,$h=1,2,\ldots,n_1$; and $v$-index is numbered along satellite moving direction,$v=1,2,\ldots,n_2$. The longitude interval for all VNs are $\Delta\Omega$ and latitude interval is $ \omega_f $. Assume the starting latitude and longitude for VN $(1,1)$ are $\phi_0$ and $\lambda_0$ respectively, then the longitude range $[\lambda_{(v,h)}^l,\lambda_{(v,h)}^h] $ can be given by 
	\begin{equation}\label{eq:VN_lon_range}
	\left\{ \begin{aligned}
	& \lambda_{(v,h)}^l=\bmod(\lambda_0 + (h-1)\Delta\Omega + 180, 360) -180,\\ 
	& \lambda_{(v,h)}^h=\bmod(\lambda_0 + h\Delta\Omega + 180, 360) -180,
	\end{aligned} \right.
	\end{equation}
	where the longitude has been normalized to $[-180^\circ,180^\circ]$. Note that $ \lambda_0 $ is changing with the relative movement between satellite orbit and the ground.
	
	The latitude range $[\phi_{(v,h)}^l,\phi_{(v,h)}^h] $ for VN $ (v,h)$ can be given by
	\begin{equation}\label{eq:VN_lat_range}
	\left\{ \begin{aligned}
	& \phi_{(v,h)}^l=90-\left| 180-\bmod \left( {\phi}'+90,360 \right) \right|, \\ 
	& \phi_{(v,h)}^h=90-\left| 180-\bmod \left( {\phi}'+\omega_f+90,360 \right) \right|,
	\end{aligned} \right.
	\end{equation}
	where the latitude has been normalized to $[-90^\circ,90^\circ]$ and
	\begin{equation}\label{eq:phi_temp}
	{\phi}'=\phi_0 + (v-1)\omega_f .
	\end{equation}
	
	Given the latitude threshold $\Phi_{P}$, {the polar regions have two latitude ranges: $[-90^\circ, -\Phi_{P}]$ and $[\Phi_{P}, 90^\circ]$.} Once the latitude of satellite satisfies $\vert \phi_s \vert >\left| \Phi_{P} \right| $, the connected H-ISLs are turned off. To simplify the VN division, $ \phi_0 $ is set to $ -\Phi_{P} $.

	\begin{figure}[t!]
		\centering
		\includegraphics[width=0.8\linewidth]{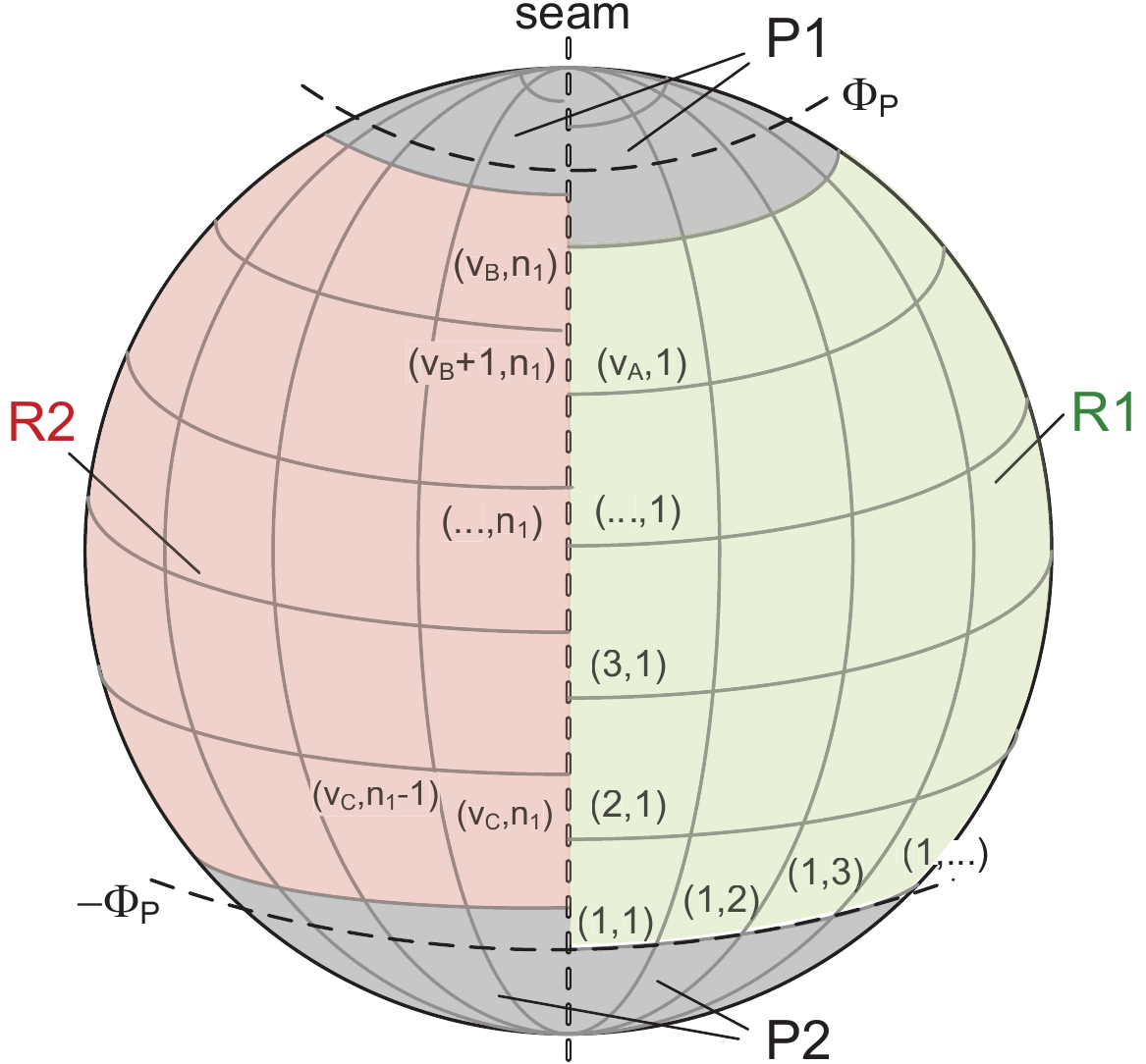}
		\caption{Celestial sphere division and VN addressing ($\Delta f =0$ )}
		\label{fig:vnspheredivision}
	\end{figure}

	\begin{table}[!t]
		\renewcommand{\arraystretch}{1}
		\caption{Four VN regions in CSD-VN}
		\label{tab:VN_regions}
		\centering
		\begin{tabular}{ccc}
			\toprule
			Region & Range of $v$-index & Inter-plane ISL state\\ 
			\midrule
			R1 & $ 1,2,...,v_A$ & on \\ 
			P1 & $ v_A+1,v_A+2,...,v_B-1$ & off \\ 
			R2 & $ v_B,v_B+1,...,v_C$ & on \\ 
			P2 & $ v_C+1,v_C+2,...,n_2$ & off \\ 
			\bottomrule 
		\end{tabular}
	\end{table}
	Based on the above-mentioned method, the celestial sphere can be partitioned into four regions and VN addresses in each region are fixed (see Fig. \ref{fig:vnspheredivision}). For each region, $h=1,2,...,n_1$ and the $ v $-indexes are listed in Table \ref{tab:VN_regions}. All H-ISLs between VNs in R1 and R2 are constantly connected while H-ISLs in P1 and P2 are constantly disconnected. Besides, VNs beside the seam have fixed $h$-index of 1 or $n_1$. The virtual link states can be completely inferred from the virtual address and keep constant. Thus, a static virtual network in the space segment is achieved by CSD-VN mapping.
	
	The virtual network is an approximate 2-D semi-Torus network and can be represented as an undirected graph $ G_V=(V,E) $, where $ V $ consists of all VNs and $ V=S $, $ E $ is the set of virtual links. $ E=E_V \bigcup E_H $, where 
	$ {{E}_{V}}=\left\{ \left. \left( \left( {{v}_{i}},{{h}_{i}} \right),\left( {{v}_{j}},{{h}_{j}} \right) \right) \right|{{h}_{i}}={{h}_{j}},\left| {{v}_{j}}-{{v}_{j}} \right|=1 \  \text{or}\  {{n}_{2}}-1 \right\}$, $ {{E}_{H}}=\left\{ \left. \left( \left( {{v}_{i}},{{h}_{i}} \right),\left( {{v}_{j}},{{h}_{j}} \right) \right) \right|\left( {{v}_{i}},{{h}_{i}} \right),\left( {{v}_{j}},{{h}_{j}} \right)\in \text{R1 or R2} \right\} $.
	
	
	According to \eqref{eq:VN_lat_range}, \eqref{eq:phi_temp} and the latitude relation between VNs and polar regions, we have 
	
	\begin{equation}\label{eq:VN_vABC1}
	\left\{\begin{array}{l}
	{v_{A} \omega_{f} \leq 2 \Phi_{P}}, \\
	{(v_{B}-1) \omega_{f} \geq \pi}, \\
	{v_{C} \omega_{f} \leq \pi+2 \Phi_{P}}.
	\end{array}\right.
	\end{equation}
	Then finally,
	\begin{equation}\label{eq:VN_vABC2}
	\left\{ \begin{array}{l}
	v_A = \lfloor n_2{\Phi_{P}}/{\pi}\rfloor,\\ 
	v_B = \lceil n_2/2 +1 \rceil,\\
	v_C =  \lfloor n_2{\Phi_{P}}/{\pi} + n_2/2\rfloor.
	\end{array} \right.
	\end{equation}
	The numbers of H-ISLs and V-ISLs are
	\begin{equation}\label{eq:ISLNumber}
	\left\{ \begin{array}{l}
	N_\text{HISL}= (n_1-1)(v_A+v_C-v_B+1),\\ 
	N_\text{VISL}= n_1 n_2,\\
	\end{array} \right.
	\end{equation}
	and the total ISLs $N_\text{ISL}= N_\text{HISL}+N_\text{VISL} $.
	
	\subsection{With inter-plane phase difference}
	When $ \Delta f \ne 0$, satellites connected by H-ISL asynchronously fly over the polar region. In both GRD-VN and CSD-VN, once a satellite enters the polar region, all H-ISLs in this row should be turned off to guarantee handover synchronization. {For satellites in the same row, the phase difference $ \Delta P $ should be minimized to achieve maximal available ISLs, and thus maximal network capacity.}
	
	In a Walker-type constellation, $\Delta f$ is specified by phasing factor $F$, $\Delta f = 2\pi F/(n_1 n_2)$, where $0 \leq F\leq n_2-1$, $F\in \mathbb{N}$.  
	Under conventional ISL connecting mode, the maximum  phase difference between satellites in a row $ \Delta P $ is up to 
	\begin{equation}\label{eq:Delta_P}
	\Delta P^0_\text{max} =(n_1 - 1)\Delta f = \frac{(n_1 -1 )F}{n_1} \omega_f .
	\end{equation}
	
	Let $ K=n_1 / F$, then $\omega_{f}=K \Delta f$. Now we revise the ISL connecting mode to minimize $\Delta P_\text{max}$. While most satellites establish forward H-ISLs (FH-ISLs) with satellite in the right plane with minimum positive $\Delta f$ (as conventional mode), we introduce the backward H-ISL (BH-ISL) connecting satellite with the nearest right neighbor with negative $\Delta f$.
	
	Let $ M(h_i) $ and $ N(h_i) $ denote the number of FH-ISLs and BH-ISLs between satellite $ (v,1) $ and $ (v,h_i) $, respectively. Then $ M(h_i)+N(h_i) = h_i-1 $. Denote the phase difference between satellites $ (v,1) $ and $ (v,h_i) $ by $ \Delta P'(h_i) $. The max value is 
	\begin{equation}\label{eq:defDeltaPMax}
	\Delta P_{\max }^{\prime}\left(h_{i}\right)=\max \left\{\Delta P^{\prime}(h) | h=1,2, \ldots, h_{i}\right\}.
	\end{equation}
	
	\textbf{Theorem 1:} When $2 < h_i \leq n_1$, $ N(h_i)= \lfloor (h_i-1)/K \rfloor $, minimum $  \Delta P'_\text{max}{(h_i)} $ is achieved and $ \Delta P'_\text{max}{(h_i)} \leq \Delta P^0_\text{max} $.

	\textbf{Proof:} 
	$ \Delta P'(h_i) = M(h_i)\Delta f + N(h_i)(\Delta f-\omega_f) \\
	=(h_i-1)\Delta f -N(h_i)\omega_f 
	= (h_i-1-N(h_i)K)\Delta f $. 
	
	Under constraint $  \Delta P'(h_i) >0 $, to minimize $  \Delta P'_\text{max}{(h_i)} $, $ N(h_i) $ satisfies $ N(h_i)= \lfloor (h_i-1)/K \rfloor $.
	Then $ \Delta P'(h_i)=\bmod(h_i-1,K) \Delta f  $, 
	$\Delta P'(h_i) $ can be written as $ \Delta P'(h_i)= \frac{(h_i-1)F-N(h_i)n_1}{F}\Delta f $. 
	
	Note that $ h_i $, $ F $, $ N(h_i) $ and $ n_1 $ are all integers, thus $\Delta P'(h_i)$ is an integral multiple of $ \Delta f /F $
	. Combining  $ \bmod(h_i-1,K)<K $, we get $ \Delta P'_\text{max}(h_i) < K\Delta f = \frac{n_1}{F} \Delta f $. Then
	$ \Delta P'_\text{max}(h_i) \leq \frac{n_1-1}{F} \Delta f  \leq \Delta P^0_\text{max} $. \hfill{$ \blacksquare $}

	To simplify issues, assume $K\in \mathbb{Z}$, then satellite $ (v,h) $ and $(v-m,h+Km)$ have the same phase. According to \textbf{Theorem 1}, when $ \Delta P'_\text{max}(h_i) $ is minimized, $ N(h_i)= \lfloor (h_i-1)/K \rfloor $, then $ N(h_i+1)-N(h_i)= \lfloor h_i/K \rfloor - \lfloor (h_i-1)/K \rfloor $,
	\begin{equation}\label{eq:Nhieq1}
	N(h_i+1)-N(h_i)= \begin{cases}
	1, & h_i=mK,m \in \mathbb{Z}, \\ 
	0, & \text{otherwise}.\\ 
	\end{cases} 
	\end{equation}
	
\begin{figure}[!t]
		\centering
		\includegraphics[width=0.9\linewidth]{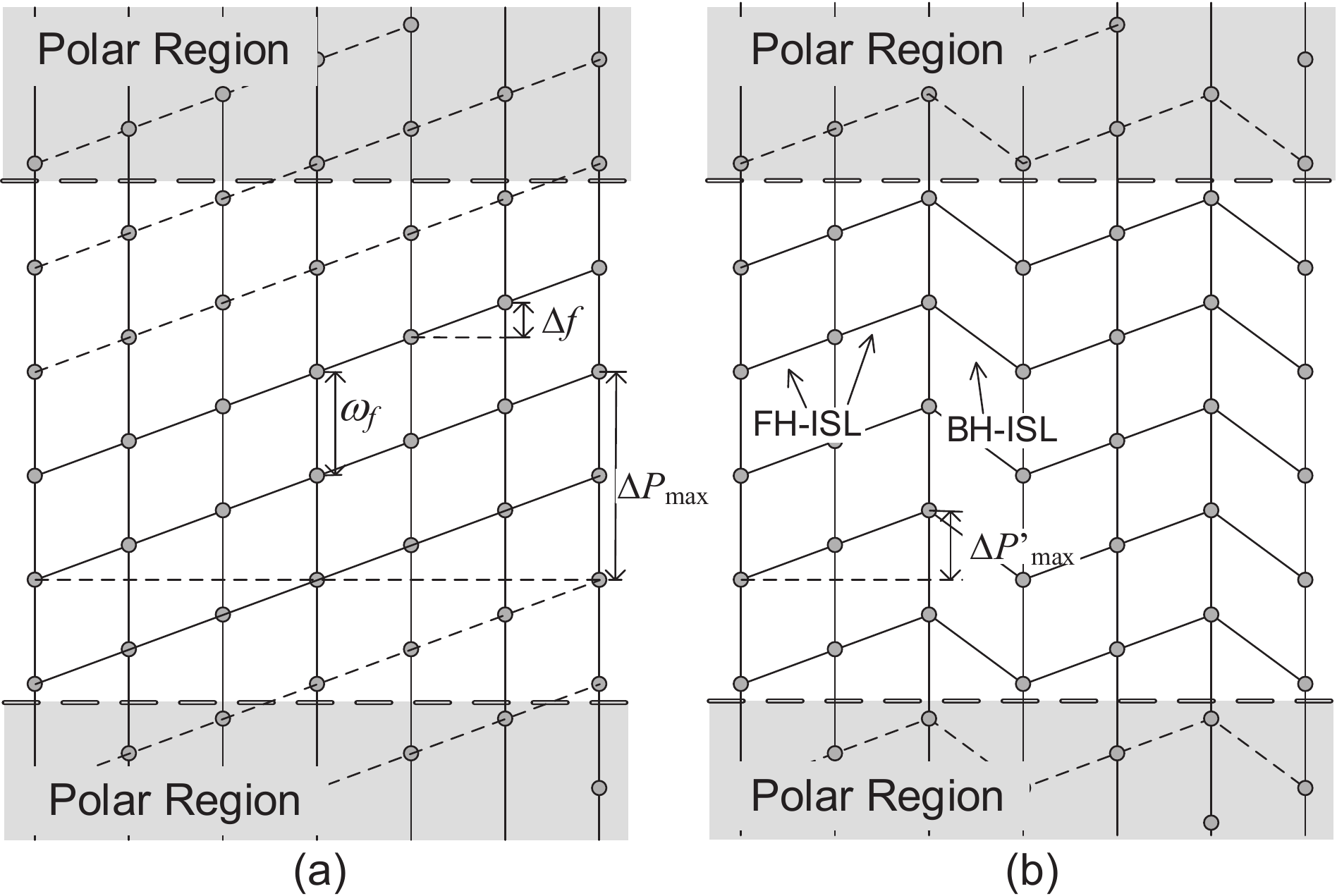}
		\caption{Comparison of inter-plane ISL connecting modes ($K$=3). (a) Conventional mode (b) Optimized mode.}
		\label{fig:ISL connecting mode}
	\end{figure}

	Based on \eqref{eq:Nhieq1}, we propose the optimized ISL connecting mode that satellites in the $mK$-th plane ($m \in \mathbb{N}$) establish BH-ISLs with the nearest satellite in the right plane, while other satellites follow the conventional mode. Fig. \ref{fig:ISL connecting mode} compares the two modes. It can be seen $ \Delta P_\text{max} $ is reduced and more ISLs become available in the optimized mode. According to \eqref{eq:defDeltaPMax} and \textbf{Theorem 1}, $ \Delta P_\text{max} $ is proved to be the minimum as 
	\begin{equation}\label{eq:Delta_P'}
	\Delta P'_\text{max} =(K - 1)\Delta f.
	\end{equation}

	When $\Delta f \ne 0$, to achieve synchronized VN switches of the whole network, under the optimized ISL connecting mode, the CSD-VN division coincides with the satellite phase. CSD-VN  can still be formulated by  \eqref{eq:VN_lon_range} and \eqref{eq:VN_lat_range}, while \eqref{eq:phi_temp} is modified to
	\begin{equation}\label{eq:phi_temp_withHPD}
	{\phi}'=\phi_0 + \bmod(h-1,K)\Delta f + (v-1)\omega_f.
	\end{equation}
	\eqref{eq:VN_vABC1} and \eqref{eq:VN_vABC2} are modified as follows
	\begin{equation}\label{eq:VN-HPD_vABC1}
	\left\{ \begin{aligned}
	& v_A  \omega_f +\Delta P'_\text{max} \leq 2\Phi_{P},\\ 
	& (v_B-1)  \omega_f  \geq \pi ,\\
	& v_C  \omega_f +\Delta P'_\text{max} \leq \pi+ 2\Phi_{P},
	\end{aligned} \right.
	\end{equation}
	\begin{equation}\label{eq:VN-HPD_vABC2}
	\left\{ \begin{aligned}
	& v_A = \lfloor n_2{\Phi_{P}}/{\pi} -(K-1)/K \rfloor,\\ 
	& v_B = \lceil n_2/2 +1 \rceil,\\
	& v_C =  \lfloor n_2{\Phi_{P}}/{\pi} + n_2/2 -(K-1)/K \rfloor.
	\end{aligned} \right.
	\end{equation}
	
\begin{figure}[t!]
		\centering
		\includegraphics[width=\linewidth]{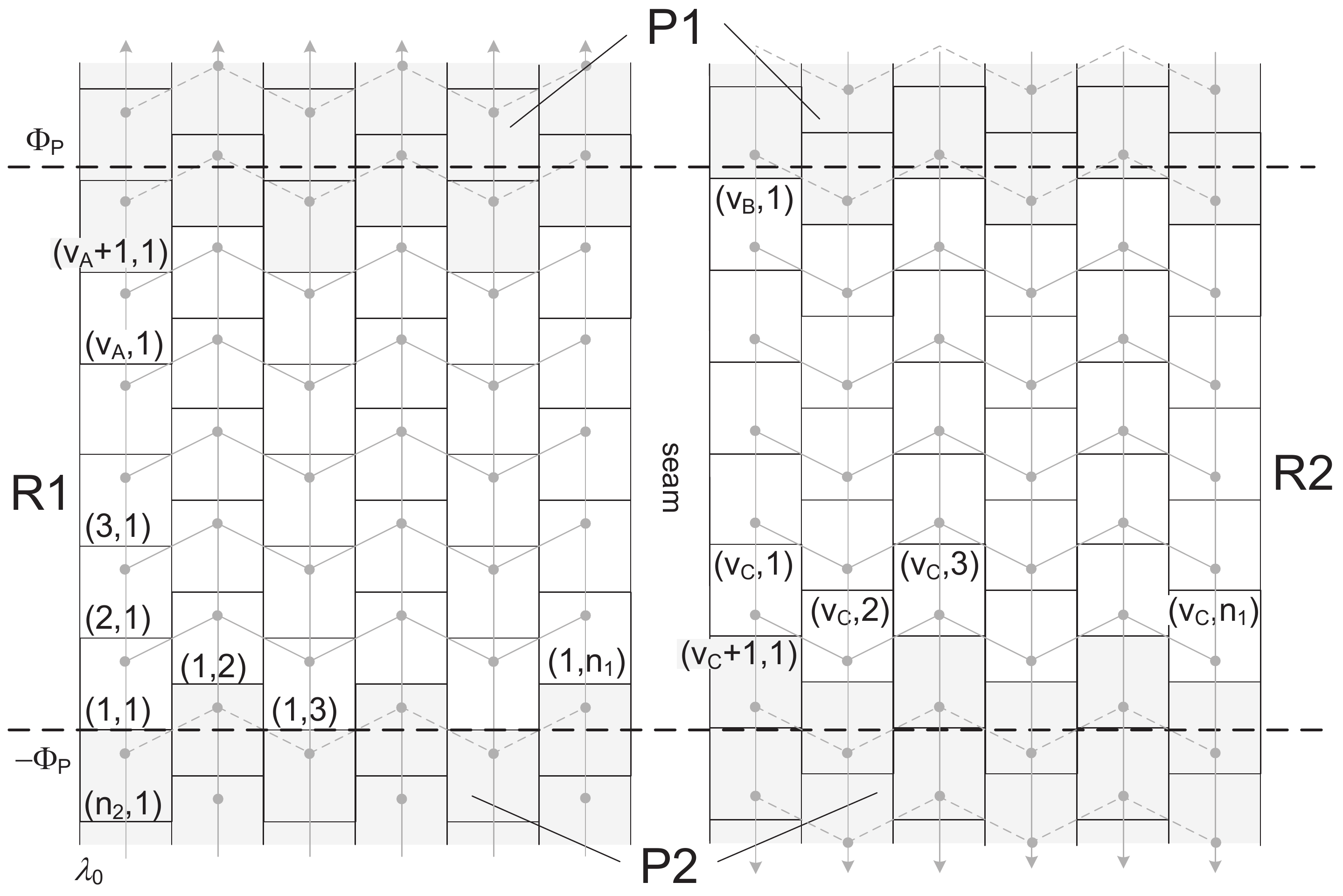}
		\caption{CSD-VN division and addressing with non-zero $ \Delta f $ ($ K=2 $)}
		\label{fig:vnHpdDivision}
	\end{figure}

	Fig. \ref{fig:vnHpdDivision} gives an example of CSD-VN division when $ K=2 $. Note that when $K \notin \mathbb{Z}$, the VN division can also be described by \eqref{eq:VN_lon_range},\eqref{eq:VN_lat_range}, \eqref{eq:phi_temp_withHPD} and \eqref{eq:VN-HPD_vABC1}.
	
	Note that with the optimized ISL mode, the network connectivity of GRD-VN can also be improved. Then The VN division should be modified to keep consistent with the satellite phase.
	
	\section{Performance comparison}
	
	\begin{table}[!t]
		\renewcommand{\arraystretch}{1.1}
		\caption{Comparison of GRD-VN and CSD-VN}
		\label{tab:Compare_Topo_shielding}
		\centering
		\resizebox{\linewidth}{!}{
			\begin{tabular}{cccc}
				\toprule 
				Topology dynamic characteristics
				& GRD-VN 1 & GRD-VN 2 & CSD-VN\\ 
				\midrule 
				Polar H-ISL disconnection & $\surd$ & $\surd$ & $\surd$ \\ 
				No H-ISL across seam & $\surd$ & $\times$ & $\surd$ \\ 
				Static topology maintenance in space& $\times$ & $\surd$ & $\surd$ \\ 
				Switch synchronization when $\Delta f \ne 0$ & $\times$ & $\times$ & $\surd$ \\ 
				Ground-to-Virtual network switching & {Partly} & {Partly} & {Partly}\\
				\bottomrule 
			\end{tabular} 
		}
		\begin{tablenotes}
			\footnotesize
			\item Note: GRD-VN 1 allows only intra-plane switches\cite{Ekici2001}, GRD-VN 2 also enables inter-plane switches\cite{Korcak2009}.
		\end{tablenotes}
	\end{table}

	\subsection{Topology dynamics}
	When earth rotation is considered, satellite adopting GRD-VN will gradually deviate from its initial VN grid. When the elevation angle $ \sigma $ decreases to zero, GRD-VN  topology cannot be maintained by satellite switches in the same orbit. While in CSD-VN, the antenna is not required to point to the fixed region, thus the topology can be maintained.
	Even if inter-plane handover is enabled to maintain the virtual topology, GRD-VN has to face new virtual topology change caused by seam movement, which is more complex in VN address updates. Moreover, in GRD-VN when $ \Delta f $ is non-zero, the asynchronous handovers within a row cause extra topology changes when satellites fly in/out polar regions. 
	
	Table \ref{tab:Compare_Topo_shielding} lists the satellite network topology dynamics and compares the performance of different VN methods. CSD-VN solves polar H-ISL on-off switches by VN division and virtual address updates. The VN division is bound with satellite moving area in inertia space so that no H-ISL across the seam and static  topology maintenance {in space} are achieved. Combining with the optimized ISL connecting mode, CSD-VN realizes synchronous switches and avoids available ISLs waste. 
	Finally, although VNs are bound with ground regions in GRD-VN, when VN network cannot be maintained, ground-to-Virtual network switching is inevitable, while in CSD-VN the switching can also be partly handled if Earth-fixed mode is enabled. An absolutely static topology for the whole system can be hardly achieved. But in general, the proposed method can shield more dynamics than conventional methods.
	
	
	\subsection{Performance of optimized ISL mode}

	\begin{figure}[t!]
		\centering
		\includegraphics[width=0.9\linewidth]{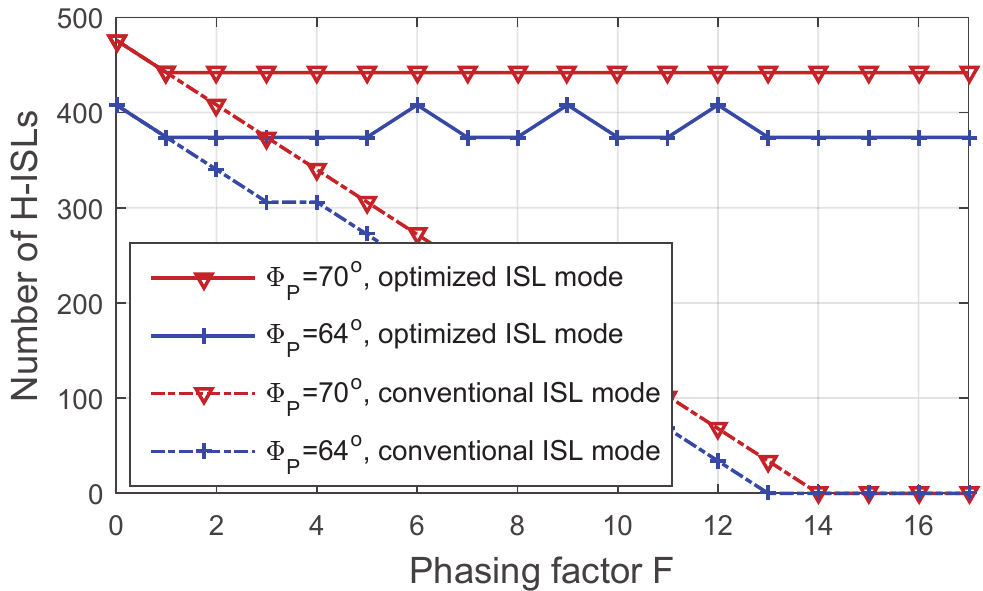}
		\caption{Available H-ISLs of VNs with different ISL modes ($ n_1 $=18, $ n_2 $=36).}
		\label{fig:numberhisl}
	\end{figure}

	The proposed optimized ISL mode can further improve the VN method by maintaining more ISLs. Fig. \ref{fig:numberhisl} compares available H-ISLs of VN methods with optimized and conventional ISL mode under different polar region ranges and $ \Delta f $. At the given constellation parameters, the proposed method maintains more available H-ISLs than conventional method owing to the optimized ISL connecting mode. With increasing $ \Delta f $, $N_\text{HISL}$ of optimized mode keeps stable and the gap expands. Fig. \ref{fig:numberhisl} also shows that $ N_\text{HISL} $ decreases with the expanding polar region. 
	
	Besides, with the optimized mode, $ \Phi_P=64^{\circ} $ case has more H-ISLs at $ F=6,9,12 $. At these points, $\lfloor (2\Phi_P- \Delta P'_\text{max})/ \omega_f \rfloor$ is larger than neighboring points. It implies that the $ \Delta f $ can be further optimized to achieve more available ISLs.

\begin{figure}[t!]
		\centering
		\includegraphics[width=0.9\linewidth]{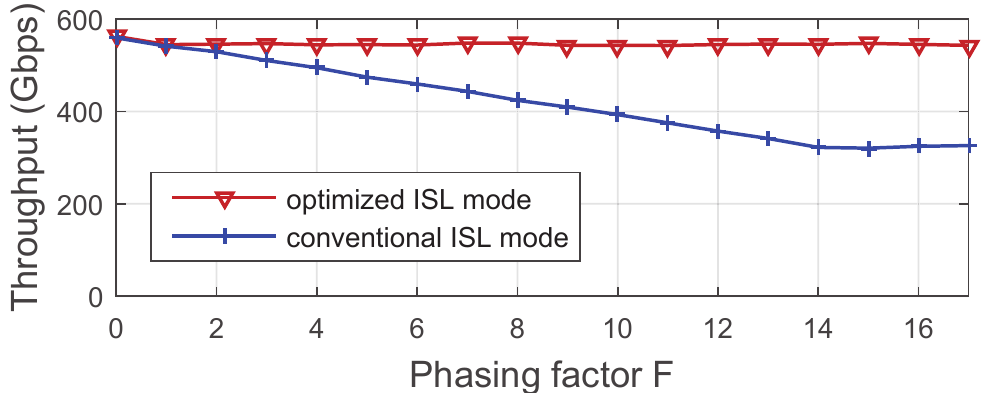}
		\caption{Throughput comparison ($ n_1=18 $, $ n_2 =36$, $\Phi_P=70^{o} $).}
		\label{fig:ComprMaxFlow}
	\end{figure}

	We compute the total system throughput using the same method in \cite{Portillo} by solving the “minimum-cost, maximum-flow” problem. The ISL capacity is set as 1 Gbps and satellite-Ground link capacity is unlimited. According to the result in Fig. \ref{fig:ComprMaxFlow}, the system throughput keeps stable with optimized ISL mode, while with conventional mode it sharply declines with the increase of $ F $ until $ F $=14, which shows a similar trend as the number of H-ISLs.
	
\begin{figure}[t!]
		\centering
		\includegraphics[width=0.9\linewidth]{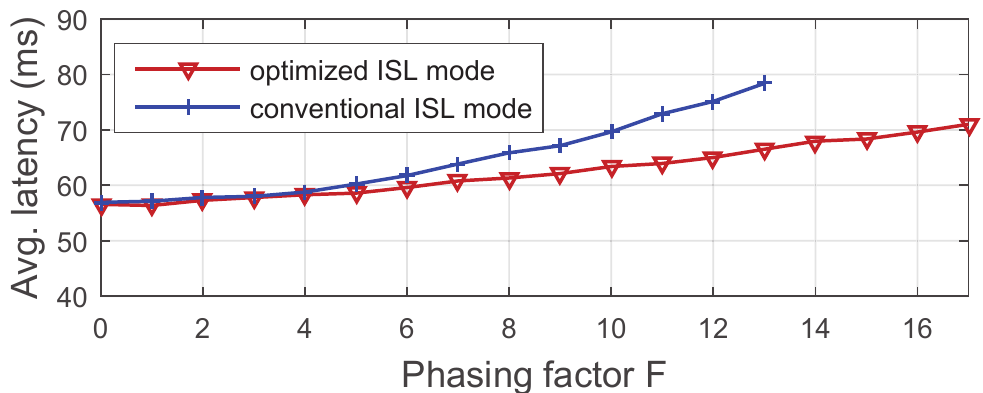}
		\caption{Latency comparison ($ n_1=18 $, $ n_2 =36 $, $\Phi_P=70^{o} $).}
		\label{fig:ComprDelay}
	\end{figure}

	Furthermore, by solving the shortest paths for 10,000 randomly generated node pairs, we obtain the average network latency  (see Fig. \ref{fig:ComprDelay}). Since the physical distance of H-ISL becomes longer with a larger $ \Delta f $, the latency will increase along with the growing $ F $. Moreover, compared to the optimized ISL mode, the latency of conventional mode grows faster. It is because the reduction of H-ISL lowers the network connectivity, packets in system with conventional ISL mode need to take a longer journey to reach the destination. In summary, the proposed VN method with optimized ISL mode is able to maintain more ISLs and keep network connectivity, thus keeping the system throughput at a high level while saving latency.

	%
	
	\section{Conclusion}
	This letter proposes a topological dynamics shielding method CSD-VN for LEO satellite networks by establishing a static virtual network in space. we formulate and optimize the VN method in both without and with phase difference scenarios. The comparison shows that, the proposed method can overcome the deficiency of the conventional VN method in seam and switching synchronization issues and also have better network performance.

	\ifCLASSOPTIONcaptionsoff
	\newpage
	\fi

	
	
	%

	\bibliographystyle{ieeetr}
	\bibliography{SatNetwork2018-VN}

	%
	%
	
	%
	
	%
	%
	%
	
	
	

\end{document}